\documentstyle[11pt,aaspp4,flushrt,epsfig]{article}
\newcommand{\al}{\alpha}

\newcommand{\La}{\Lambda}
\newcommand{\Om}{\Omega}

\newcommand{\ra}{\rightarrow}

\newcommand{\vs}{{v_{{\scriptscriptstyle 12}}}}
\newcommand{\xl}{{\xi^{{\scriptscriptstyle (1)}}}} 
\newcommand{\xxl}{{\xi_{{\scriptscriptstyle 1}}}} 
\newcommand{\xq}{{\xi^{{\scriptscriptstyle (2)}}}} 
\newcommand{\xxq}{{\xi_{{\scriptscriptstyle 2}}}} 
\newcommand{\se}{{\sigma_{{\scriptscriptstyle 8}}}} 
\newcommand{\be}{\begin{equation}}
\newcommand{\ee}{\end{equation}}
\newcommand{\gsim}{\stackrel{>}{\sim}}
\newcommand{\lsim}{\stackrel{<}{\sim}}
\newcommand{\bea}{\begin{eqnarray}}
\newcommand{\eea}{\end{eqnarray}}
\newcommand{\bean}{\begin{eqnarray*}}
\newcommand{\eean}{\end{eqnarray*}}
\newcommand{\dd}{\partial}
\newcommand{\xb}{\bar{\xi}}
\newcommand{\xbb}{\bar{\hspace{-0.08cm}\bar{\xi}}}
\newcommand{\xbl}{{\xb^{{\scriptscriptstyle (1)}}}} 
\newcommand{\xbq}{{\xb^{{\scriptscriptstyle (2)}}}}

\newcommand{\RoA}{{{\rho}_{{\scriptscriptstyle A}}}}

\newcommand{\rone}{{{\vec r}_{{\scriptscriptstyle 1}}}}

\newcommand{\rtwo}{{{\vec r}_{{\scriptscriptstyle 2}}}}
\newcommand{\rA}{{{\vec r}_{{\scriptscriptstyle A}}}}
\newcommand{\va}{{{\vec v}_{{\scriptscriptstyle A}}}}
\newcommand{\vv}{{{\vec v}_{{\scriptscriptstyle 12}}}}
\newcommand{\vone}{{{\vec v}_{{\scriptscriptstyle 1}}}}
\newcommand{\vtwo}{{{\vec v}_{{\scriptscriptstyle 2}}}}
\newcommand{\xig}{{\xi_{{\scriptscriptstyle {\rm g}}}}}
\newcommand{\vsg}{{v_{{\scriptscriptstyle 12{\rm g}}}}}
\newcommand{\da}{{{\delta}_{{\scriptscriptstyle A}}}}
\newcommand{\done}{{{\delta}_{{\scriptscriptstyle 1}}}}
\newcommand{\dtwo}{{{\delta}_{{\scriptscriptstyle 2}}}}
\newcommand{\dg}{{\delta_{{\scriptscriptstyle {\rm g}}}}}
\newcommand{\dgA}{{\delta_{{\scriptscriptstyle {\rm g}A}}}}
\lefthead{Juszkiewicz, Durrer, \& Springel }
\righthead{Mean pairwise velocity}

\begin{document}
\tighten
\title{Dynamics of pairwise motions} 
\author{Roman Juszkiewicz\altaffilmark{1,3}, 
Volker Springel\altaffilmark{2},
and Ruth Durrer\altaffilmark{1}}

\altaffiltext{1}{D{\'e}partement de Physique Th{\'e}orique, Universit{\'e}
de Gen{\`e}ve, CH-1211 Gen{\`e}ve, Switzerland}
\altaffiltext{2}
{Max-Planck-Institut f{\"u}r Astrophysik, D-85740 Garching, Germany}
\altaffiltext{3}
{On leave from
Copernicus Astronomical Center, 00-716 Warsaw, Poland}

\begin{abstract} 
We derive a simple closed-form expression,
relating $\vs(r)$ -- the mean relative velocity of pairs of galaxies
at fixed separation $r$ -- 
to the two-point correlation function  of mass density
fluctuations, $\xi(r)$. We compare our analytic model for $\vs(r)$
with N-body simulations, 
and find excellent agreement in the entire dynamical range
probed by the simulations ($0.1 \lsim \xi \lsim 1000$).
Our results can be used to estimate the cosmological
density parameter, $\Om$, 
directly from redshift-distance surveys, like Mark III.
\end{abstract}


\tighten

\section{An analytical model for $\vs(r)$}

Most dynamical estimates of the cosmological density parameter, $\Omega$,
use the gravitational effect of departures from a strictly homogeneous
distribution of mass. One such dynamical estimator can be constructed
by using an equation expressing the conservation of particle pairs
in a self-gravitating gas.  This equation was derived by
\cite{md77} from the BBGKY theory (see also \cite{jp80},
hereafter LSS). Consider a pair of particles
at a comoving separation vector ${\vec x}$ and cosmological time $t$,
moving with a 
mean (pair-weighted) relative velocity 
$\vs(x,t) \, {\vec x}\,/x$. It's magnitude
is related to the two-point correlation function
of density fluctuations, $~\xi(x,t)~$, by the
pair conservation equation (LSS), 
\be
 {a\over 3[1+\xi(x,a)]} \, {\dd\xb(x,a)\over \dd a} \; = \; - \, 
{\vs(x,a)\over Hr}~, \label{full}
\ee
where $a(t)$ is the expansion factor, $r = ax$ is the proper separation,
$H(a)$ is the Hubble parameter,
while $\xb(x,a)$ is the two-point correlation function, averaged over
a ball of comoving radius $x$ :
\be
\xb(x,a) \; = \; 3 x^{-3} \, \int_0^x \xi(y,a)y^2 {\rm d}y \; .
\ee
At the present cosmological time $a = 1$, $x = r$ and 
$H = 100~h$km s$^{-1}$Mpc$^{-1}$, where $h$ is the conventional dimensionless
parameterization for the Hubble constant. 
There are two well known approximate
solutions of~(\ref{full}). They 
are: the small separation limit, where
$\xi(x) \gg 1$ (stable clustering regime), and
the large separation limit, where $|\xi| \ll 1$ (linear regime).
The stable clustering solution is (LSS, \S  71)
\be
\vs(x,a) \; = \;  - \, Hr \; ,
\ee 
as expected for virialized clusters on sufficiently small scales.
The linear solution is given by the first terms in
perturbative expansions for $\vs$ and $\xi$, which
for the correlation is given by 
\be
\xi \; = \; \xl \, +  \, \xq \, + \ldots \; ,
\ee
with $\xq = O\left[ \xl \right]^2$, etc.
The growing mode of the linear solution is 
\be
\xl(x,a)  \; =  \; \xxl(x)D(a)^2 \; ,
\ee
where $D(a)$ is the usual linear growth factor (LSS, \S 11;
we neglect the decaying mode).
The general technique for deriving $\xq$ and higher order terms  
for initially Gaussian density fluctuations in an Einstein-de Sitter
universe was introduced by Juszkiewicz et al. (1980, 1984),
Vishniac (1983) and Fry (1984); their results were recently generalized
to a wider class of cosmological models including those
considered below (\cite{rj93,fb95}). These calculations
show that the second order term in the expansion for
$\xi$ can be written as 
\be
\xq(x,a) \; =  \; \xxq(x) \, D(a)^4 \; ,
\ee
where $\xxq$ is a function of $x$ alone.
Substituting  $\, \xi = D^2 \,\xxl \, + \, D^4 \, \xxq \,$ 
into eq.~(\ref{full}) and solving
for $\vs$ we get
\be
\vs \; = \; -\, {\textstyle{2\over3}} \, Hrf \,
\left[\, \xbl \; - \; \xbl \xl \; + \; 2\,\xbq \,
\right] \; + \; O\left[\xl\right]^3 \; , 
\label{2ndvs}
\ee
where $f \equiv d \ln D / d \ln a \approx \Omega^{0.6}$ 
(which is a good approximation if $\Lambda = 0$ or 
$\Om+\Om_{\Lambda}=1$; $\Om_{\Lambda} \equiv \Lambda/3H^2$).
The bars above $\xi$-s denote averaging  
over a ball of comoving radius $x$. If the logarithmic slope
of the correlation function,
$\gamma(x) \, \equiv \, - \,d \ln\xl(x,a) /d \ln x$, 
is a constant in the range $0 \le \gamma < 2$. Then
$\xbq$ is related to $\xbl$ by a simple closed-form expression,    
\be
\xbq(x,a) \; =  \; \alpha(\gamma) \, \left[\xbl(x,a)\right]^2 \; ,
\label{parabolic}
\ee
where $\alpha$ is a constant (\cite{el96}), and
the function $\alpha(\gamma)$ can be expressed in terms
of Euler's $\Gamma$ functions (\cite{rs96}). 
In the range $0 \le \gamma < 2$, this 
expression is well approximated by a fitting formula
\be
\alpha \; = \; 1.84 \, - \, 1.1 \,\gamma \,
- \, 0.84\,(\gamma/2)^{10} \; .
\label{alpha}
\ee
For $\gamma \ge 2$, perturbation theory diverges,
but N-body simulations suggest that the 
relation (\ref{parabolic}) remains valid, provided
the $\alpha(\gamma)$ dependence for $\gamma \gsim 1.5$
is derived from
the so-called universal scaling relation (USH) --
an empirical formula for the nonlinear power spectrum,
extracted from $\gamma \ge 1$ scale-free and
cold dark matter (CDM) simulations (\cite{jmw95,rs96}).
Since the USH approach fails when $\gamma < 1$, while the 
perturbation theory (PT)
fails for for $\gamma \ge 2$, we propose to trade
accuracy for an extended range of validity and interpolate 
the expression $\alpha(\gamma)$ between the PT formula for
$\gamma < 1$ and the USH result for $\gamma > 1$ (\cite{rs96},
Fig.20), and replace eq.~(\ref{alpha}) with
\be
\alpha \; \approx \; 1.2 \, - \, 0.65 \, \gamma \; .
\label{volker}
\ee 
The allowed range of slopes for the above expression is 
$0 < \gamma < 3$.
To extend the range of validity of our perturbative solution 
(\ref{2ndvs}) into the nonlinear
regime, we propose the following Ansatz:
\be 
\vs(x,a)  \; = \; 
- \, {\textstyle {2\over 3}}\,Hrf\,
\xbb(x,a)\,\left[\,1 \; + \;\al\;\xbb(x,a) \; \right]\; ,
\label{2nd} \; 
\ee
\be 
\xbb(x,a) \; = \; \xb(x,a) \, [1 + \xi(x,a)]^{-1} \; .
\ee
The expression~(\ref{2nd}) agrees  with the perturbative
expansion~(\ref{2ndvs}) when $\xi \rightarrow 0$; it 
also approximates the stable clustering limit when
$x \ra 0$. Indeed, when $\xi$ is large, 
$\;
- \vs/Hr \; \approx \; (2/3)f(\Omega)(1+\alpha)\;$ which is
of order unity for all reasonable values of $\Om$ and $\alpha$.

Eq.~(\ref{2nd}) can be used to predict the relative velocity
$\vs$ at any separation $r$. The only input necessary for that is
the correlation function in an interval of separations $\leq r$.
This is very different from the semi-analytic expression for $\vs(r)$,
derived by Mo at al. (1997). 
In their case, the knowledge of the present $\xi(r)$
is not sufficient to calculate $\vs(r)$; they use the USH approach
and a large array of parameters instead.

\section{N-body simulations}

In this section we compare our analytic expression for $\vs$
with the results from high-resolution AP$^3$M simulations 
of 256$^3$ dark matter particles, kindly provided to us by the
Virgo collaboration (\cite{aj98}). 
We consider four members of the  
CDM family: an open model (OCDM), a zero curvature low-$\Om$
model ($\Lambda$CDM), and two models with an Einstein-de Sitter
metric -- the `standard CDM' model (SCDM)
and its modified version ($\tau$CDM).
Following Jenkins et al., the values, assigned
to parameters  $(h, \,\Om,\,\Om_{\Lambda}, \,
\se$) are: (0.7, 0.3, 0, 0.85) for OCDM, (0.7, 0.3, 0.7, 0.9)
for $\Lambda$CDM, and (0.5, 1, 0, 0.6) for the SCDM and
$\tau$CDM which has
extra large-scale power (added in an ad hoc manner,
described by \cite{aj98}). Here and below $\se$ is the 
rms dark matter density contrast in a ball
of radius 8 $h^{-1}$Mpc.

Since CDM-like models are not scale-free, eq.~(\ref{volker}) does
not apply. In principle, we should therefore
calculate $\alpha(x) \equiv \xbq(x,a)/[\xbl(x,a)]^2$ 
for each considered power spectrum and each separation $x$, 
using standard perturbative techniques
(\cite{el96,rs96}). However, as we will show below, these
calculations can be significantly simplified by finding an effective slope,
$\gamma_{\rm eff}$, which provides
`best fit' $\alpha$ and $\vs$ when substituted in
eq.~(\ref{volker}) and (\ref{2nd}). The precise
value of $\alpha$ is unimportant in the stable clustering
regime as well as in the linear regime, when the term
quadratic in $\xbb$ is sub-dominant. Hence, the precision in
$\alpha(x)$ matters only at the
boundary between the linear and nonlinear regimes,
say at $\xi(x,t) = 1$. One of the possible definitions
of the effective slope is therefore given by
\be
\gamma_{\rm eff} \; = \; - (d \ln \xi /d \ln x)|_{\xi = 1} \; .
\label{gamma}
\ee
One can also choose $\gamma_{\rm eff}$ as follows.
In Fig.~1, we plot logarithmic slopes of $\xl(x)$
and $\xi(x)$ (the latter is measured from the simulations).
Both curves agree at large separations as they should, 
apart from small differences arising from noise in the measurement (we
use only a finite number of bins and pairs to measure $\xi$), and from
finite box-size and cosmic-variance effects (\cite{aj98}).
However, there is a well-defined scale (marked with an asterisk)
at which the non-linear slope turns away from the linear theory
prediction, marking the onset of the non-linear regime.
We take $\gamma_{\rm eff}$ to be the logarithmic slope 
of $\xbl$ at that scale. The resulting slopes are:
$\gamma_{\rm eff}=1.67$ (SCDM), $1.46$
($\Lambda$CDM), $1.49$ (OCDM), and $1.40$ ($\tau$CDM). 
The alternative definition (\ref{gamma})
gives, respectively: $\gamma_{\rm eff}=1.67$, $1.47$,
$1.45$, and $1.28$. The advantage of the former definition is that 
it is more closely related to
observations because it uses $\xi(x)$ only rather
than $\xi$ along with $\xl(x)$.
The linear correlation function is easy to calculate under
the controlled conditions of an N-body experiment
but it can not be easily determined from observations.

In Figure~2 we test our Ansatz (\ref{2nd}) against
N-body measurements. For comparison, we also plot three
other approximations for $\vs(r)$, considered earlier in
the literature:
\be
{\rm A)} \;\;\; \vs = - {\textstyle {2\over 3}} Hrf  
\xbl \;\;\; , \qquad {\rm B)} \;\;\; 
\vs =  - {\textstyle {2\over 3}} Hrf  
\xb \;\;  , \qquad {\rm C)}\;\;\;
\vs =  - {\textstyle {2\over 3}} Hrf \,  
\xbb \;\; 
\; .
\label{vslin}
\ee
(A) and (B) are two variants of linear theory predictions, and 
(C) is an improvement over
linear theory, suggested by Peebles (LSS, \S 71).
Figure 2 shows that the deviations from linear theory are small
at large separations, as they should.
The range of validity of the Peebles formula (C) is 
already considerably wider than that of linear theory. However, 
our new nsatz provides by far the best approximation. 
In fact, it covers the entire dynamical range!

The scale at which the linear approximation becomes acceptable
depends on the amplitude of fluctuations; it increases
with increasing $\se$. For example, for a power-law correlation
function, we have
\be
\xb(r) \; = \; \se^2 \, F(\gamma)\, r^{-\gamma} \; ,
\qquad {\rm where}
\ee
\be
F(\gamma) \; = \; (16 \, h^{-1}{\rm Mpc})^{\gamma}
(4-\gamma)(6-\gamma)/24 \; ,
\ee
and eq.~(\ref{vslin}~B) can be rewritten as
\be
\vs(r) \, \approx \, - {\textstyle {2\over 3}} \,\se^2\Om^{0.6}
\, H\, F(\gamma) \, r^{1-\gamma} \, \approx \,
- \, 667 \,\se^2 \, \Om^{0.6} \; {\rm km/s} \; ,
\label{vs-se}
\ee
where the expression after the last ``$\approx$'' sign assumes
$\gamma = 1.75$ and $r = 10\,h^{-1}$Mpc. The 
relative error in the latter expression,
introduced by linear theory, can be calculated from
eqs. (\ref{2nd}) and (\ref{volker}). It depends on $\se$ only; for
$\se = 1$  and 0.6, linear theory overestimates $\, |\vs| \,$ by
24\% and 10\%, respectively.

\section{Velocity bias}

So far we considered the dynamics of pairwise motions of
dark matter particles. However, for practical applications,
it is necessary to understand the relation between $\vs(r)$
and the relative pairwise velocity of galaxies, $\vsg(r)$.  
We define the galaxy clustering bias as the square root of
the ratio between the galaxy and the dark matter correlation functions:
$ \; b(r,t)^2 = \, \xig(r,t)/\xi(r,t)$. 

Observations suggest that there is no velocity bias:
splitting the Mark~III catalogue~(\cite{will}) into subsamples of
elliptical and spiral galaxies has no effect on $\, \vsg \;$
(\cite{rj98}). Similar results follow from recent numerical simulations,
which account for dissipative processes, important
for galaxy formation. These models
do show some {\it clustering bias} on small scales; however,
there is no {\it velocity bias}, and $\, \vs = \vsg \;$ (\cite{gk98}).

As we show now, this contradicts the  simplest  toy model
for bias, where $b \approx 1/\se$ is a constant.  
In this prescription, also known as linear bias,
the galaxy density contrast at position
$\rA$ is simply given by $\dgA \approx b\da$, where
$\da \; \equiv \; \RoA / \langle \rho \rangle - 1$ is the mass density
fluctuation, and $A = 1,2, \ldots$ enumerate galaxy positions.
In the fluid approach, the mean pairwise
velocity between two points at distance $r$ is given by
\be
\vv(r)  = 
{{ \langle(\vone - \vtwo ) (1 + \done)(1 +\dtwo ) \rangle}
	\over {1  + \xi(r)}}~ ,
\label{def}
\ee
where $\va$ is the peculiar velocity 
at a point $\rA$, $r=|\rone - \rtwo|$ is the separation, and $\xi(r) =
\langle \done \dtwo \rangle$. For the galaxy pair
density-weighted relative velocity, $\vsg$, the matter density
field in the above expression, $\delta$, has to be replaced by
$\dg$. In the limit of large separations $(\delta, \, \xi \ra 0)$, the linear
biasing model, applied to eq.~(\ref{def}),
gives $\vsg(r) = b\vs(r)$, and since
$\vs \propto \se^2\Om^{0.6}$ one obtains $\vsg \propto
\se\Om^{0.6}$ (Fisher et al. 1994). On small scales, where
$1+\delta\sim \delta$ and $1+\xi\sim \xi$, the factors of $b$ 
in the denominator and numerator cancel
and $\vsg(r) = \vs(r)$. 
This unphysical behavior shows the limitations
of the linear biasing model. Similar results
can be obtained from considerations, based on the
continuity equation and gravitational clustering.
Gravity tends to remove bias and
$b$ has to evolve with time and separation~(\cite{jf96}).
The results of Kauffmann et al. (1998) fit neatly into this picture:
their measurements of $b(r)$ at 
the `present time' (redshift = 0) are significantly different from
unity only on small scales where $\xi \gg 1$ and gravity is no longer  
the only force determining the dynamics; at large scales,
however, gravity dominates and $b =1$. Since the dependence on $b$
in eq.~(\ref{def}) cancels out when $\xi \gg 1$, it is
not surprising that the simulations give $\, \vs = \vsg \,$ on
all scales. Summarizing, we conclude that a realistic
model of biasing should lead to unbiased pairwise velocities;
the linear bias model disagrees with observations as well as
with simple theoretical considerations.

\section{Conclusions}

The main result of this paper is the expression~(\ref{2nd}) -- 
an approximate analytic solution of the pair conservation equation.
Our equations~(\ref{volker}) - (\ref{gamma})
reproduce the results of numerical simulations on all
scales from the strongly non-linear to the linear regime. 

Our results can be used to estimate $\Om$ from redshift-distance
surveys, like Mark III (\cite{pf98,rj98}).
On very large scales,  $\vs(r)$ is proportional to 
$\Om^{0.6}\se^2$. On intermediate scales (the mildly
non-linear regime) this degeneracy is removed and $\Om$ and $\se$ can
be measured separately because the $\alpha \; \xbb$ term in 
eq.~(\ref{2nd}) is $\Om$-independent. 
These properties make our estimator
complementary to other estimators.
Indeed, the {\small POTENT}\
method~(\cite{dekel}) and the cluster abundances~(\cite{nb98,eke98})
are sensitive to $\beta \equiv \Om^{0.6}\se$; 
the supernovae~(\cite{snIa,snIb})
distances  measure $q_o \equiv (\Om/2)-\Om_\La$;  
finally, the position of 
acoustic peaks in the CMB power spectrum~(\cite{DZS}) is
sensitive to $\Om+\Om_\La$. The advantage of our estimator over 
the CMB peaks method is model-independence; its
advantage over {\small POTENT}
is simplicity~(\cite{pf98}). 

Our results can also be used to study the nature of biasing and
to test the gravitational instability theory. Indeed, one can
use the galaxy correlation function, $\xi_g(r)$, estimated
from galaxy redshift or angular surveys to predict $\vs(r)$
which could then be compared to a $\vs(r)$ estimated from
redshift-distance surveys (\cite{eg98}).

\acknowledgements

We thank Antonaldo Diaferio, Enrique Gazta{\~n}aga and
Roman Scoccimarro for insightful comments. This work was supported by
grants from the Polish Government (KBN grants 
No. 2.P03D.008.13 and 2.P03D.004.13),
the Tomalla Foundation, by the Poland-US
M. Sk{\l}odowska-Curie Fund and by the Swiss National Science Foundation.
RJ thanks Simon White for his hospitality at MPA in Garching.

\begin{figure}
\begin{center}
\resizebox{8cm}{!}{\includegraphics{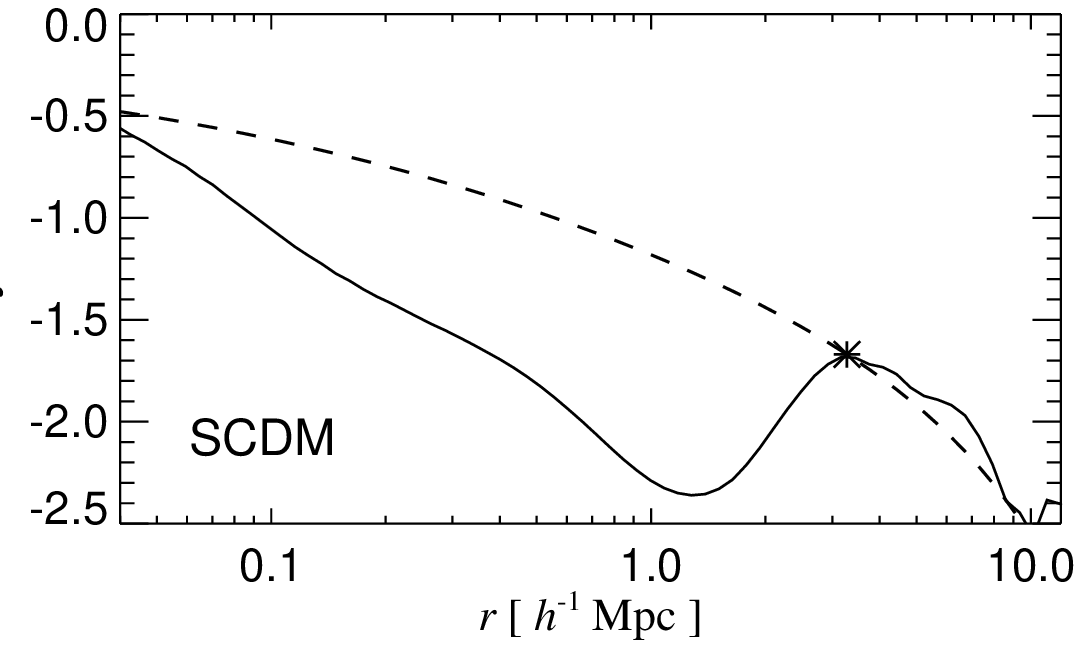}}
\resizebox{8cm}{!}{\includegraphics{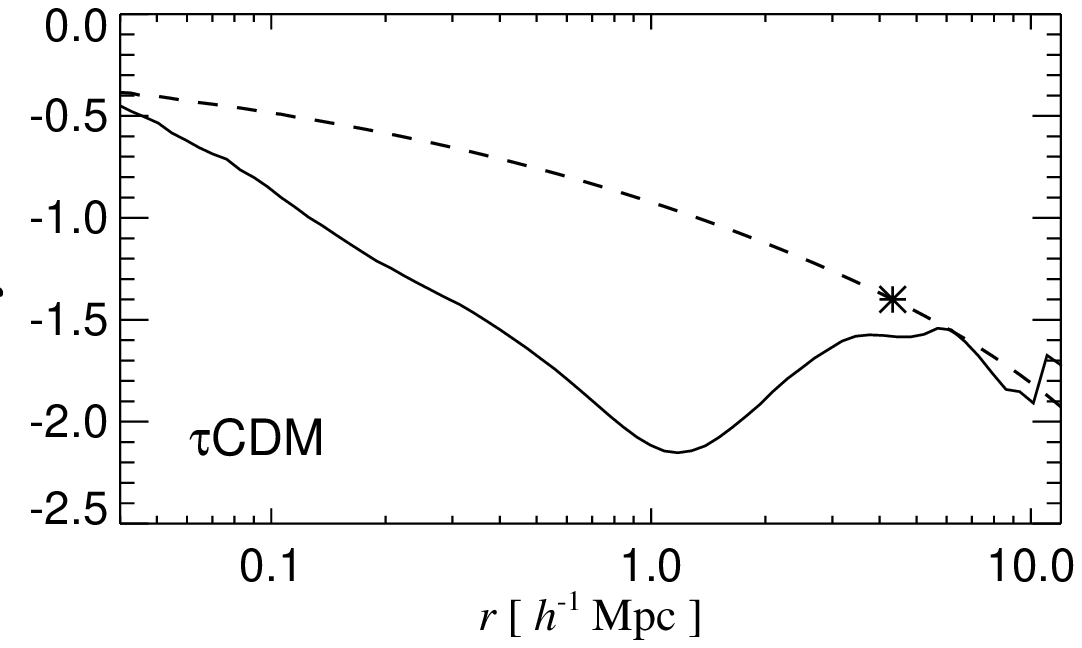}}
\resizebox{8cm}{!}{\includegraphics{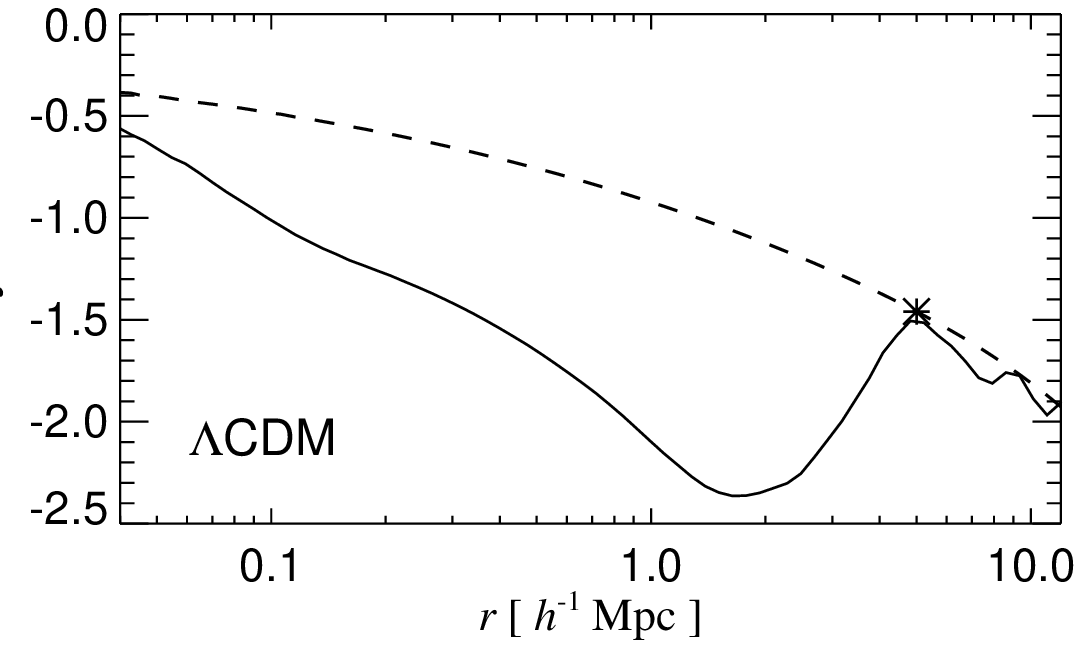}}
\resizebox{8cm}{!}{\includegraphics{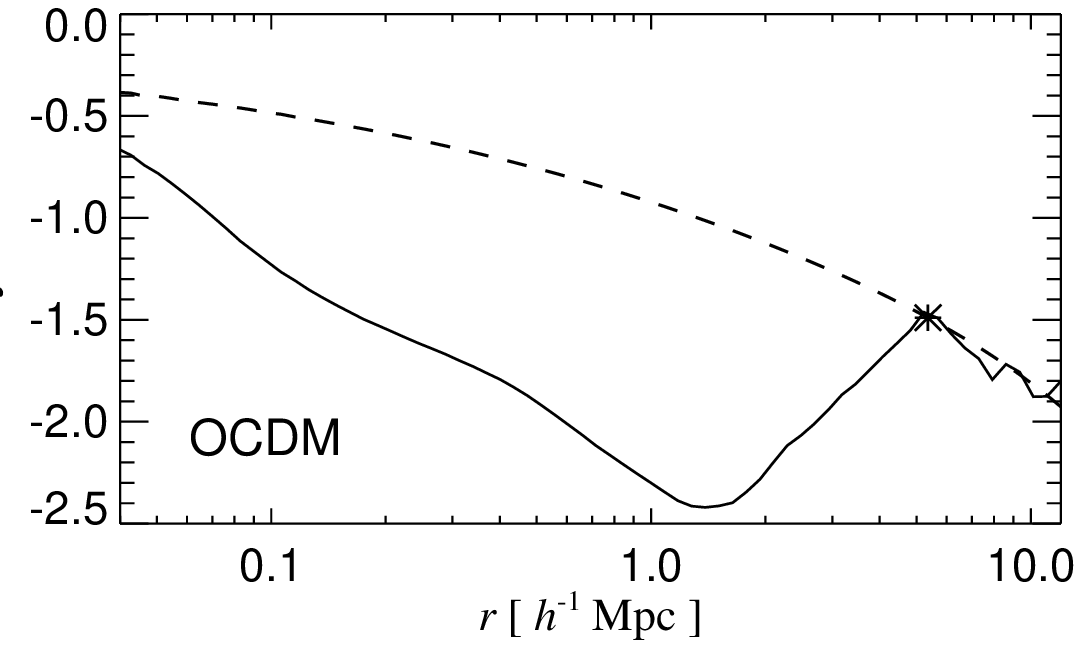}}
\caption
{Logarithmic slope $d\ln\xi/ d \ln  r$   
of the linear theory correlation function (dashed), and 
the measured non-linear correlation function (solid) for the four Virgo
simulations which we have analyzed. 
The asterisks mark the effective slopes $\gamma_{\rm eff}$ 
used in equations (\ref{volker}) and (\ref{2nd}), respectively.
\label{fig1}}
\end{center}
\end{figure}

\begin{figure}
\begin{center}
\resizebox{8cm}{!}{\includegraphics{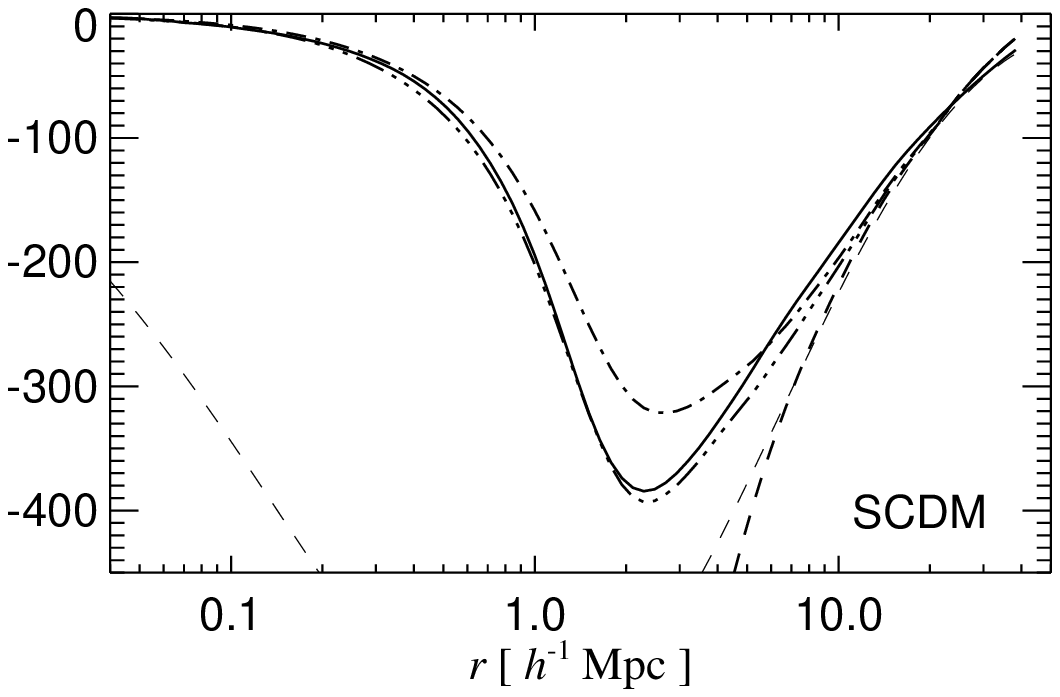}}
\resizebox{8cm}{!}{\includegraphics{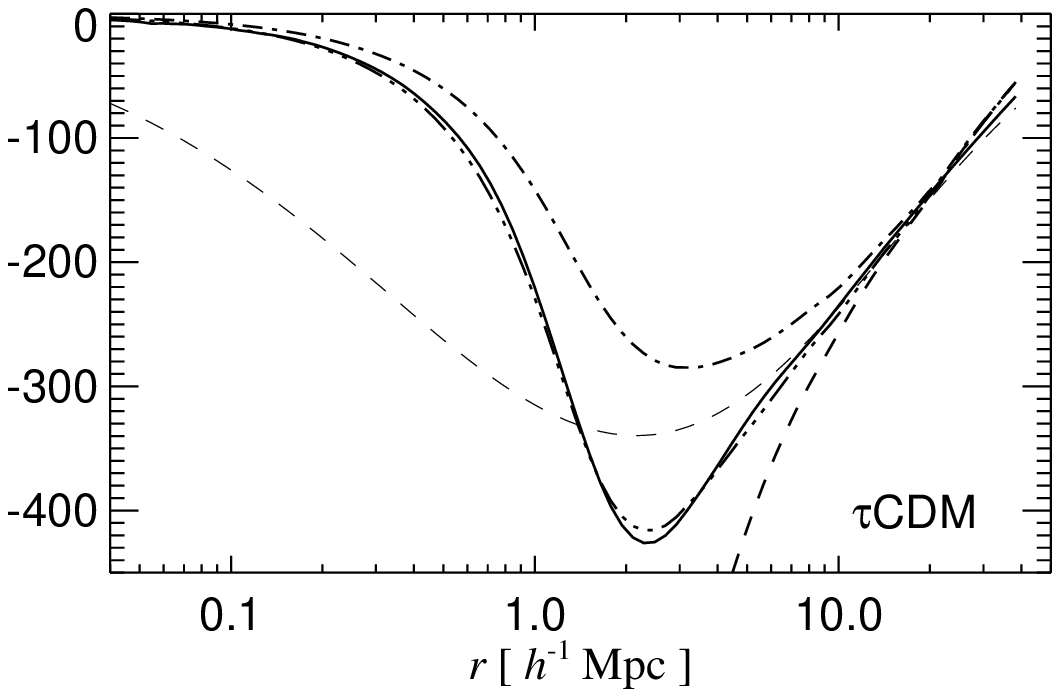}}
\resizebox{8cm}{!}{\includegraphics{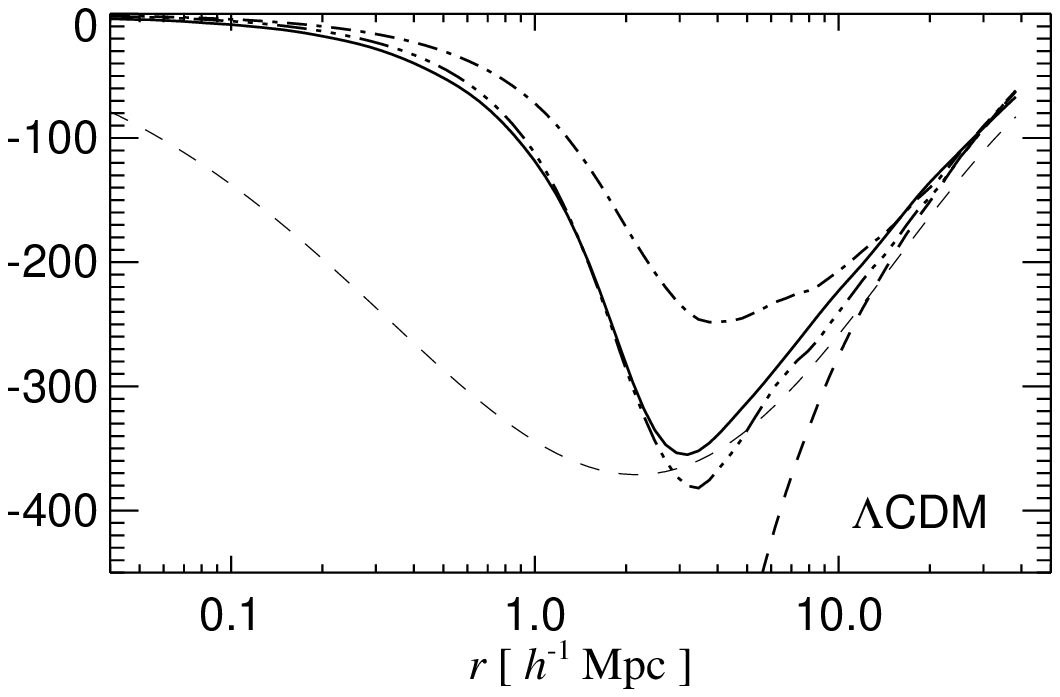}}
\resizebox{8cm}{!}{\includegraphics{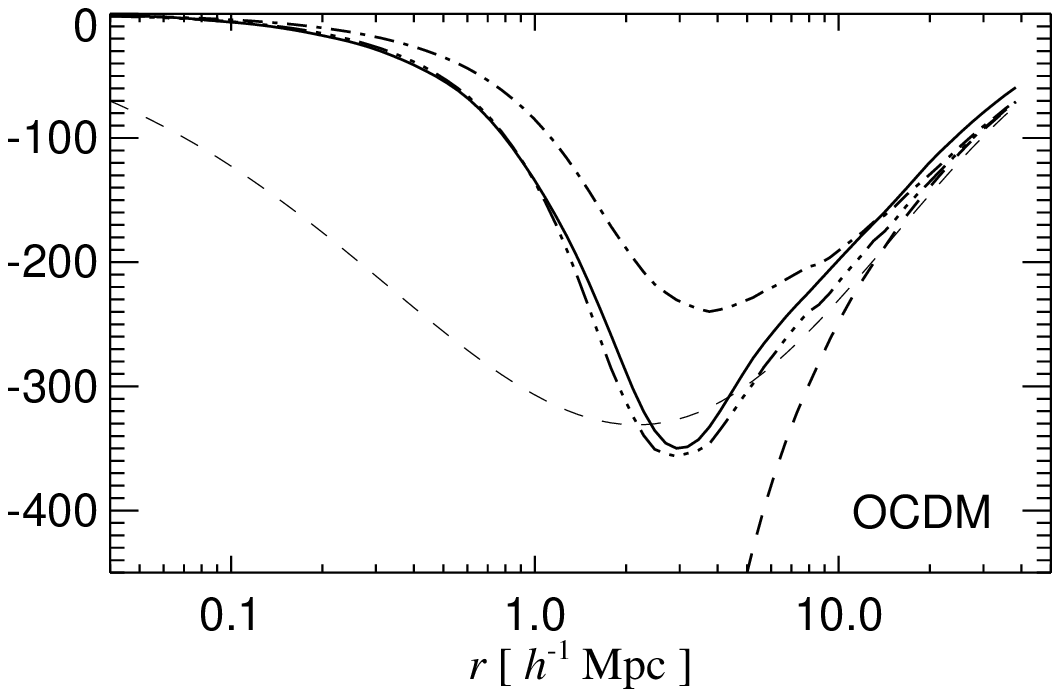}}
\caption
{The mean pairwise velocity $v_{12}$ of the Virgo simulations (solid
lines) compared 
with four closed-form approximations to solutions of the pair 
conservation equation: the two versions of the linear approximation, 
eq.~\ref{vslin} (A), and (B), are plotted as thin and thick
dashed curves respectively; the Peebles approximation, eq.~\ref{vslin}
(C), is shown as dot-dashed curve; and eq.~(\ref{2nd}) -- the Ansatz
proposed in this paper -- is drawn as dot-dot-dot-dashed line.
\label{fig2}}
\end{center}
\end{figure}

\end{document}